\documentclass[12pt]{article}
\begin{document}


\null\vskip-24pt \hfill KL-TH/02-05  \vskip0.3truecm
\begin{center}
\vskip 3truecm {\Large\bf
On  marginal deformation of WZNW model and PP-wave
limit of deformed $AdS_{3}\times S^{3}$ string geometry}\\
\vskip 1.5truecm
{\large\bf Ruben Manvelyan
 \footnote{On leave from Yerevan Physics Istitute, email:{\tt
manvel@physik.uni-kl.de, manvel@moon.yerphi.am} }
}\\
\vskip 1truecm
{\it Department of Physics, Theoretical Physics\\
University of Kaiserslautern, Postfach 3049 \\
67653 Kaiserslautern, Germany}\\

\end{center}
\vskip 3truecm \centerline{\bf Abstract}

We discuss the Penrose limit of the classical string geometry
obtained from a truly marginal deformation of $SL(2)\otimes SU(2)$
WZNW model.
\newpage
The PP-wave geometry \cite{Penrose,BFHP} and exactly solvable
string theory in the corresponding backgrounds
\cite{Met,MetTset,RTset} have received growing attention after
appearance of the article of BMN \cite{BMN}, in which the authors
proposed some holographic description. This new holographic
description was based on the fact that PP-waves can be obtained as
a Penrose limit of different supersymmetric $AdS_{p}\times
S^{10/11-p}$ (p=5/7,4) geometries of IIB and M-theories and
therefore inherit the holographic properties of the latter. On the
other hand, strings in PP-waves can be quantized exactly
\cite{Met}, as in  flat space at least in the light-cone gauge,
and thus in this case  of the string/gauge theory dualities one
can work with the full string spectrum.

Another example of the exactly quantized and holographic string is
the string in the $AdS\times S^{3}\times M^{4}$ background
\cite{GKSg, KSg, GK, AGS, SW1}, the exact quantization of which is
connected with the well known equivalence between the string in
 the $AdS_{3}\times S^{3}$ background and the $SL(2,R)\otimes SU(2)$
Wess-Zumino-Novikov-Witten model, and the holographic conjecture
allows to construct the boundary conformal algebra from the string
world sheet objects. The Penrose limit of the $AdS_{3}\times
S^{3}$ string, quantization in this case and some holographic
properties were investigated recently in \cite{dav, mathur,
ppcft}.

In this note we study a deformed $AdS_{3}\times S^{3}$ background
using a truly marginal deformation of the WZNW model
\cite{Schwarz} for the $SL(2)\otimes SU(2)$ group manifold given
by the operator $\int d^{2}z
(J_{sl}\bar{J}_{su}+J_{sl}\bar{J}_{su})$ which is bilinear in
corresponding $U(1)$ currents of the Cartan subalgebra. This type
of deformation for the $SU(2)$ and $SU(2)\otimes SU(2)$ WZNW
models was considered in \cite{Sen},\cite{GivKir}, and in \cite{F}
for the case of $SL(2)$ in the null direction. The main idea of
this consideration at that time was to use $O(d,d)$ transformation
of the compactified string (WZNW model) to generate a new
classical solution of string theory from known ones, because this
transformation can be obtained from the marginally deformed WZNW
model and therefore will maintain world sheet  conformal
invariance. Some deformation of the $SL(2)\times SU(2)$ WZNW model
was used in \cite{CvetTset} for the description of the
five-dimensional non-extreme rotating black holes with large
charges. We will consider a truly marginal deformation of the
$SL(2)\otimes SU(2)$ WZNW model using the method  of direct
solution of the corresponding differential equation for deformed
action and currents developed in \cite{GivKir} for the $SU(2)$
case. Then we consider the Penrose limit of this deformed string
geometry  and find the corresponding PP-wave background.

There are different parameterizations of the WZNW model for the
group manifolds $SL(2,R)/AdS_{3}$ and $SU(2)/S^{3}$ :
\begin{equation}\label{WZ}
  S(g)=\frac{k}{32\pi}\int_{\partial B}Tr\left(\partial_{\mu}
  g^{-1}\bar{\partial}^{\mu}g\right)
  +\frac{k}{48\pi}\int_{B}\varepsilon^{ijk}
  Tr\left(\partial_{i}gg^{-1}\partial_{j}gg^{-1}\partial_{k}gg^{-1}\right)
\end{equation}
First of all we can use Euler angles for both models:
\begin{equation}\label{EA}
  g=e^{i\theta_{1}\sigma_{3}}e^{ix\sigma_{2}}e^{i\theta_{2}\sigma_{3}}
\end{equation}
In these coordinates the action (\ref{WZ}) for for $SU(2)$ is:
\begin{equation}\label{SU}
  S[x,\theta]=\frac{k}{2\pi}\int d^{2}z \left(\partial x
  \bar{\partial}x+\partial\theta_{1}
  \bar{\partial}\theta_{1}+\partial\theta_{2}\bar{\partial}\theta_{2}+
  2\cos{2x} \partial\theta_{2}\bar{\partial}\theta_{1}\right).
\end{equation}
The action for $SL(2,R)$ is obtained\footnote{ Another useful
parameterization in the case of $SL(2,R)$ only is the Gauss
decomposition parameterization:
\begin{eqnarray}\label{gd}
g= \left(\begin{array}{cc}
   1 & 0 \\
   \gamma & 1
 \end{array}\right)\left(\begin{array}{cc}
   e^{y}& 0 \\
   0 & e^{-y}
 \end{array}\right)\left(\begin{array}{cc}
   1 & \bar{\gamma} \\
   0 & 1
 \end{array}\right).\nonumber
\end{eqnarray}
This decomposition leads to the action
\begin{eqnarray}\label{wz2}
  S[y,\gamma]=\frac{k}{2\pi}\int d^{2}z \left(\partial y
  \bar{\partial}y+e^{2y}\partial\bar{\gamma}\bar{\partial}\gamma\right).
\nonumber\end{eqnarray}} from (\ref{SU}) by replacing $x$ by $iy$
and $k$ by $-k$:
\begin{equation}\label{SL}
  S[y,\psi]=\frac{k}{2\pi}\int d^{2}z \left(\partial y
  \bar{\partial}y-\partial\psi_{1}
  \bar{\partial}\psi_{1}-\partial\psi_{2}\bar{\partial}\psi_{2}-
  2\cosh{2y} \partial\psi_{2}\bar{\partial}\psi_{1}\right)
\end{equation}

The $AdS_{3}\times S^{3}$ string equipped with the NS-NS
antisymmetric field $B_{\mu \nu }$ and constant dilaton
$\varphi_{0} $ is just the combination of these two Lagrangians.
The forms of the metric and $B$-field
\begin{eqnarray}\label{met}
  &&ds^{2}=k\left(-\cosh^{2}y dt^2 + dy^{2} + \sinh^{2}y d\psi^{2} +
  \cos^{2}x d\tau^2 + dx^2 + \sin^{2}x d\theta^{2}\right),\,\,\,\,\,\,\,\,\\
  &&H=dB=2k\sinh2ydy\wedge dt \wedge d\psi +2k\sin2x dx \wedge
  d\tau \wedge d\theta
\end{eqnarray}
can be read off from the action
\begin{eqnarray}\label{SLSU}
S^{AdS^{3}\times S^{3}}[x,y,\psi ,\theta]&=&\frac{k}{2\pi}\int
d^{2}z \left(\partial x
  \bar{\partial}x+\partial\theta_{1}
  \bar{\partial}\theta_{1}+\partial\theta_{2}\bar{\partial}\theta_{2}+
  2\cos{2x} \partial\theta_{2}\bar{\partial}\theta_{1}\right.\nonumber\\
  &&\left.+\partial y
  \bar{\partial}y-\partial\psi_{1}
  \bar{\partial}\psi_{1}-\partial\psi_{2}\bar{\partial}\psi_{2}-
  2\cosh{2y} \partial\psi_{2}\bar{\partial}\psi_{1}\right)
\end{eqnarray}
using symmetrization and antisymmetrization  of the fields and
the following change of variables
\begin{eqnarray}\label{change}
\theta_{1,2}=\frac{\tau\pm\theta}{2},\qquad
\psi_{1,2}=\frac{t\pm\psi}{2}
\end{eqnarray}
We return to this coordinate system later after deformation of our
theory, but for the moment we  use expression (\ref{SLSU}) for the
$AdS^{3}\times S^{3}$ string because this form of the WZNW action
is useful in the following respect. The equations of motion
corresponding to the variations of the fields
$\psi_{i},\theta_{i}$ lead directly to the conservation of the
following chiral currents
\begin{eqnarray}
 J_{\theta}&=&\partial\theta_{1} +\cos{2x}\partial\theta_{2} ,\label{c1}\\
 \bar{J}_{\theta}&=&\bar{\partial}\theta_{2}+\cos{2x}\bar{\partial}\theta_{1} ,\label{c2}\\
J_{\psi}&=&-\partial\psi_{1}-\cosh{2y}\partial\psi_{2} ,\label{c3}\\
\bar{J}_{\psi}&=&-\bar{\partial}\psi_{2}-\cosh{2y}\bar{\partial}\psi_{1}\label{c4}
\end{eqnarray}
We now consider a minimal deformation with deformation parameter
$\lambda$ of fields and interactions, preserving chiral conserved
currents in the following way:
\begin{eqnarray}
  S^{SL(2)\times SU(2)}_{\lambda}&=&\frac{k}{2\pi}\int
d^{2}z \left(\partial x
  \bar{\partial}x+\partial\theta^{\lambda}_{1}
  \bar{\partial}\theta^{\lambda}_{1}+\partial\theta^{\lambda}_{2}\bar{\partial}\theta^{\lambda}_{2}+
  2\Sigma^{\lambda}_{\theta\theta}(x,y) \partial\theta^{\lambda}_{2}\bar{\partial}\theta^{\lambda}_{1}\right.\nonumber\\
  &&\left. +\partial y\bar{\partial}y-\partial\psi^{\lambda}_{1}
  \bar{\partial}\psi^{\lambda}_{1}-\partial\psi^{\lambda}_{2}\bar{\partial}\psi^{\lambda}_{2}-
  2\Sigma^{\lambda}_{\psi\psi}(x,y) \partial\psi^{\lambda}_{2}\bar{\partial}\psi^{\lambda}_{1}\right.\nonumber\\
  &&\left.+2\Sigma^{\lambda}_{\theta\psi}(x,y) \partial\theta^{\lambda}_{2}\bar{\partial}\psi^{\lambda}_{1}+
  2\Sigma^{\lambda}_{\psi\theta}(x,y) \partial\psi^{\lambda}_{2}\bar{\partial}\theta^{\lambda}_{1}\right)\label{lag}
\end{eqnarray}
This form of deformed interactions preserves the existence of the
conserved chiral currents
\begin{eqnarray}
J_{\theta}^{\lambda}&=&\partial\theta^{\lambda}_{1}
+\Sigma^{\lambda}_{\theta\theta}(x,y)
\partial\theta^{\lambda}_{2}+\Sigma^{\lambda}_{\psi\theta}(x,y)
\partial\psi^{\lambda}_{2} ,\label{C1}\\
 \bar{J}_{\theta}^{\lambda}&=&\bar{\partial}\theta^{\lambda}_{2}+\Sigma^{\lambda}_{\theta\theta}(x,y)
 \bar{\partial}\theta^{\lambda}_{2}+\Sigma^{\lambda}_{\theta\psi}(x,y)
 \bar{\partial}\psi^{\lambda}_{1} ,\label{C2}\\
J_{\psi}^{\lambda}&=&-\partial\psi^{\lambda}_{1}-\Sigma^{\lambda}_{\psi\psi}(x,y)\partial\psi^{\lambda}_{2}
+\Sigma^{\lambda}_{\theta\psi}(x,y)\partial\theta^{\lambda}_{2} ,\label{C3}\\
\bar{J}_{\psi}^{\lambda}&=&-\bar{\partial}\psi^{\lambda}_{2}-\Sigma^{\lambda}_{\psi\psi}(x,y)\bar{\partial}\psi^{\lambda}_{1}
+\Sigma^{\lambda}_{\psi\theta}(x,y)\bar{\partial}\theta^{\lambda}_{1}.\label{C4}
\end{eqnarray}
Here all $\Sigma$ s are unknown functions of $\lambda$ and radial
parameters $x,y$, and $\theta^{\lambda}_{i}$ and
$\psi^{\lambda}_{i}$ are initial fields $\theta_{i}$ and
$\psi_{i}$ mixed with some general $\lambda$-dependent $4\times 4$
matrices ${\cal A}(\lambda)$:
\begin{eqnarray}\label{ab}
  \left(%
\begin{array}{c}
\psi^{\lambda}_{1}\\
\psi^{\lambda}_{2}\\
\theta^{\lambda}_{1} \\
  \theta^{\lambda}_{2} \\
\end{array}%
\right)&=& {\cal A}(\lambda)\left(%
\begin{array}{c}
\psi_{1} \\
\psi_{2}\\
\theta_{1}\\
\theta_{2}\\
\end{array}%
\right)\\
\partial_{\lambda} \left(%
\begin{array}{c}
  \psi^{\lambda}_{1}\\
\psi^{\lambda}_{2}\\
\theta^{\lambda}_{1} \\
  \theta^{\lambda}_{2} \\
\end{array}%
\right)&=&\partial_{\lambda}{\cal A}(\lambda){\cal A}^{-1}(\lambda)\left(%
\begin{array}{c}
  \psi^{\lambda}_{1}\\
\psi^{\lambda}_{2}\\
\theta^{\lambda}_{1} \\
  \theta^{\lambda}_{2} \\
\end{array}%
\right) ,
\end{eqnarray}

We will be interested in the mixed deformation preserving zero
value of the beta function. This corresponds to a truly marginal
deformation which is bilinear in chiral conserved currents from
the Cartan subalgebra of the $SL(2)$ and $SU(2)$ algebras. For
obtaining the truly marginal deformation we have to solve the
equation
\begin{equation}\label{main}
  \partial_{\lambda}S^{SL(2)\times
  SU(2)}_{\lambda}=\frac{2k}{2\pi}\int d^{2}z\left(J_{\theta}^{\lambda}
  \bar{J}_{\psi}^{\lambda}+
  J_{\psi}^{\lambda}\bar{J}_{\theta}^{\lambda}\right)
\end{equation}
with the corresponding initial conditions
\def\openone{\leavevmode\hbox{\small1\kern-3.8pt\normalsize1}}
\begin{eqnarray}
  \Sigma^{\lambda=0}_{\theta\theta}(x,y)&=& \cos{2x} ,\nonumber\\
  \Sigma^{\lambda=0}_{\psi\psi}(x,y)&=& \cosh{2y} ,\label{cond}\\
\Sigma^{\lambda=0}_{\theta\psi}(x,y)&=&\Sigma^{\lambda=0}_{\psi\theta}(x,y)=0 ,\nonumber\\
{\cal A}(0)=\openone .\label{conda}
\end{eqnarray}
Solving (\ref{main}) we have to fix  all unknown $\Sigma$ s and
${\cal A, B}$ matrices also. The substitution of (\ref{lag}) in
(\ref{main}) leads to the following set of differential equations
for unknown $\Sigma$ terms
\begin{eqnarray}\label{difur1}
  \partial_{\lambda}\Sigma_{\theta\theta}&=&
  \Sigma_{\theta\theta}\left(\Sigma_{\psi\theta}+\Sigma_{\theta\psi}\right),\\
\partial_{\lambda}\Sigma_{\psi\psi}&=&
\Sigma_{\psi\psi}\left(\Sigma_{\psi\theta}+\Sigma_{\theta\psi}\right)\label{difur2},\\
\partial_{\lambda}\Sigma_{\psi\theta}&=&
\Sigma_{\psi\theta}^2-\Sigma_{\theta\theta}\Sigma_{\psi\psi}+1\label{difur3},\\
\partial_{\lambda}\Sigma_{\theta\psi}&=&
\Sigma_{\theta\psi}^2-\Sigma_{\theta\theta}\Sigma_{\psi\psi}+1\label{difur4}
,
\end{eqnarray}
with the initial conditions (\ref{cond}) and  the following
equations for $\theta_{i}, \psi_{i}$ with the initial condition
(\ref{conda}):
\begin{eqnarray}\label{dif1}
  &&\partial_{\lambda}\psi_{1}^{\lambda}=\theta_{2}^{\lambda},\qquad
  \partial\psi_{2}^{\lambda}=\theta_{1}^{\lambda},\nonumber\\
  &&\partial_{\lambda}\theta_{1}^{\lambda}=-\psi_{2}^{\lambda},\quad
  \partial_{\lambda}\theta_{2}^{\lambda}=-\psi_{1}^{\lambda},\\
  &&\psi_{i}^{\lambda=0}=\psi_{i},\qquad
  \theta_{i}^{\lambda=0}=\theta_{i}.
\end{eqnarray}
The solution of the latter is :
\begin{eqnarray}\label{amat}
  \psi_{1}^{\lambda}&=&\psi_{1}\cos\lambda +\theta_{2}\sin\lambda ,\\
\psi_{2}^{\lambda}&=&\psi_{2}\cos\lambda +\theta_{1}\sin\lambda ,\\
\theta_{1}^{\lambda}&=&\theta_{1}\cos\lambda -\psi_{2}\sin\lambda ,\\
\theta_{2}^{\lambda}&=&\theta_{2}\cos\lambda -\psi_{1}\sin\lambda
,
\end{eqnarray}
which is just a rotation in the space of
$(\psi_{1},\psi_{2},\theta_{1},\theta_{2})$ coordinates.

Then from the system (\ref{difur1}) to (\ref{difur4}) we can
easily derive the condition
$\Sigma^{\lambda}_{\psi\theta}=\Sigma^{\lambda}_{\theta\psi}$ and
obtain a solvable equation,
\begin{eqnarray}
  \Sigma^{\lambda}_{\theta\theta}&=&C \Sigma^{\lambda}_{\psi\psi} ,
  \qquad C=\frac{\cos{2x}}{\cosh{2y}}\label{r1}   ,\\
  \partial_{\lambda}M(\lambda)&=&2\Sigma^{\lambda}_{\psi\theta} ,
   \qquad M(\lambda)=\ln{\Sigma^{\lambda}_{\psi\psi}}\label{r2} ,\\
\frac{1}{2}\partial^{2}_{\lambda}M &-&
\frac{1}{4}(\partial_{\lambda}M)^{2}-C \exp(2M) -1=0 .\label{r3}
\end{eqnarray}
For solving (\ref{r3}) we can define the new system
\begin{eqnarray}
  F(M)&=&\partial_{\lambda}M(\lambda) ,
  \qquad  \partial^{2}_{\lambda}M =
  \frac{1}{2}\partial_{M}(F^{2}) ,\label{f1}\\
\frac{1}{4}\partial_{M}(F^{2}\exp(-M))&=&\exp(-M)-C\exp(M)
.\label{f2}
\end{eqnarray}
With solutions for $F$ and $M$
\begin{eqnarray}
  F(M)&=&2\sqrt{B\exp(M)-C\exp(2M)-1} ,\label{f}\\
  \lambda+D&=&\frac{1}{2}\arctan\left[\frac{B\exp{M}-2}{2\sqrt{B\exp{M}-C\exp{2M}-1}}\right] .\label{m}
\end{eqnarray}
Here $B, D$ are the arbitrary $\lambda$-independent functions .

Finally using the initial conditions (\ref{cond}) we obtain the
solutions
\begin{eqnarray}\label{sol}
  \Sigma^{\lambda}_{\psi\psi}(x,y)&=&
\frac{\cosh{2y}}{\cos^{2}{\lambda}+\sin^{2}\lambda\cosh{2y}\cos{2x}} ,\\
\Sigma^{\lambda}_{\theta\theta}(x,y)&=&
\frac{\cos{2x}}{\cos^{2}{\lambda}+\sin^{2}\lambda\cosh{2y}\cos{2x}} ,\\
\Sigma^{\lambda}_{\psi\theta}(x,y)&=&
\frac{\sin{\lambda}\cos{\lambda}(1-\cos{2x}\cosh{2y})}
{\cos^{2}{\lambda}+\sin^{2}\lambda\cosh{2y}\cos{2x}} .
\end{eqnarray}
From this solution and from the action (\ref{lag}), comparing with
the sigma model expressions
\begin{eqnarray}\label{sigma}
S&=&\frac{k}{2\pi}\int
dz^{2}\left(G_{\mu\nu}(\lambda)+B_{\mu\nu}(\lambda)\right)\bar{\partial}X^{i}\partial X^{j} ,\\
\phi(\lambda)&=&\phi_{0}+\frac{1}{2}\ln{\frac{\det{G(0)}}{\det{G(\lambda)}}}
,
\end{eqnarray}
we can find the metric of the new deformed string solution (in
coordinates rotated with the angle $\lambda$), $H=dB$ and dilaton
fields:
\begin{eqnarray}\label{newsol}
&&ds^{2}=k\left(dx^{2}+dy^{2}-d\psi_{1}^2-d\psi^{2}_{2}+d\theta_{1}^{2}+d\theta_{2}^{2}
+\frac{2}{\Delta}\left[\cos{2x}d\theta_{1}d\theta_{2}\right.\right.\nonumber\\
&&\left.\left.-\cosh{2y}d\psi_{1}d\psi_{2}
+\sin{\lambda}\cos{\lambda}(1-\cos{2x}\cosh{2y})
(d\psi_{2}d\theta_{1}+d\psi_{1}d\theta_{2})\right] \right) ,\,\,\,\,\\
&&H=\frac{2k}{\Delta^{2}}\left[2\cos^{2}\lambda\sinh2y dy\wedge
d\psi_{1}\wedge d\psi_{2}+2\sin^{2}\lambda\cosh^{2}2y\sin2x
dx\wedge d\psi_{1}\wedge d\psi_{2}\right.\nonumber\,\,\,\,\\
&&+2\cos^{2}\lambda\sin2x dx\wedge d\theta_{1}\wedge
d\theta_{2}+2\sin^{2}\lambda\cos^{2}2x\sinh2y dy\wedge
d\theta_{1}\wedge d\theta_{2}\nonumber\,\,\,\,\\
&&\cos{2x}\sinh{2y}\sin{2\lambda}dx\wedge\left(d\psi_{1}
\wedge d\theta_{2}+d\theta_{1}\wedge d\psi_{2}\right)\nonumber\,\,\,\,\\
&&\left.-\cosh{2y}\sin{2x}\sin{2\lambda}dy\wedge\left(d\psi_{1}
\wedge d\theta_{2}+d\theta_{1}\wedge d\psi_{2}\right)\right] ,\,\,\,\,\\
&&\phi(\lambda)=\phi_{0}+\ln{\Delta},\,\,\,\,\qquad\Delta=
\cos^{2}{\lambda}+\sin^{2}\lambda\cosh{2y}\cos{2x} .
\end{eqnarray}
We will be interesed in the Penrose limit of this solution. For
obtaining this we have to find the appropriate coordinates near
the null geodesic. We can find this limit using the following
change of the coordinate system:
\begin{eqnarray}\label{change1}
  &&x=\frac{r}{\sqrt{k}},\qquad\qquad y=\frac{\rho}{\sqrt{k}} ,\\
  &&\psi_{1}=\frac{\mu x^{+}+\psi\cos\lambda -\theta\sin\lambda
  }{2},\nonumber\\ &&\psi_{2}=\frac{\mu x^{+}-\psi\cos\lambda
  +\theta\sin\lambda}{2} ,\\
  &&\theta_{1}=\frac{\mu x^{+}-(k\mu)^{-1}x^{-}+\theta\cos\lambda+\psi\sin\lambda}{2},\,\,
\nonumber\\&&\theta_{2}=\frac{\mu
x^{+}-(k\mu)^{-1}x^{-}-\theta\cos\lambda-\psi\sin\lambda}{2}
,\qquad
\end{eqnarray}
and take $k\rightarrow \infty $. As a result we obtain the
following PP-wave solutions for supergravity fields:
\begin{eqnarray}\label{ppm}
&&ds^{2}=-2dx^{+}dx^{-}-\mu^{2}\left(r^{2}+\rho^{2}-
\sin2\lambda\left[r^{2}-\rho^{2}\right]\right)dx^{+2}+
d{\vec{\rho}}^{\,\,2}+d{\vec{r}}^{\,\,2} ,\\
&&H=4\mu(\cos\lambda+\sin\lambda)\rho d\rho\wedge dx^{+} \wedge
d\psi+4\mu (\cos\lambda-\sin\lambda)r dr \wedge
  dx^{+} \wedge d\theta ,\qquad\\
  &&\phi(\lambda)=\phi_{0},\qquad
  d\vec{\rho}^{\,\,2}=d\rho^{2}+\rho^{2}d\psi^{2},\quad
  d{\vec{r}}^{\,\,2}=dr^{2}+r^{2}d\theta^{2}.
\end{eqnarray}
Therefore we obtain in the Penrose limit again the quadratic
metric and constant fluxes but now we have some additional
deformation parameter $\lambda$ leading to different masses and
fluxes
\begin{equation}\label{mpm}
  \mu_{\pm}^{2}=\mu^{2}\left(1\pm
  \sin2\lambda\right)=\mu^{2}(\cos\lambda\pm\sin\lambda)^{2}
\end{equation}
for transversal modes of the corresponding light-cone string lying
in two dimensional planes coming from  $AdS_{3}$ and from $S^{3}$
parts of the initial manifold. Indeed  in the point $\lambda=0$ we
obtain the usual PP-wave Penrose limit of $AdS_{3}\times S^{3} $
but for the $\lambda=\pm\frac{\pi}{4}$ we have separation of the
flat 2d Euclidian space, i.e.:
\begin{eqnarray}\label{spec}
&&ds^{2}=-2dx^{+}dx^{-}-2\mu^{2}\rho^{2}dx^{+2}+d{\vec{\rho}}^{\,\,2}+d{\vec{r}}^{\,\,2} ,\\
&&H=4\sqrt{2}\mu\rho d\rho\wedge dx^{+} \wedge d\psi .
\end{eqnarray}
Thus we can deduce that the Penrose limit of our deformed string
geometry  connects the two PP-wave plane times flat plane points
with the usual 6d PP-wave coming from the $AdS_{3}\times S^{3}$
geometry. It is clear that this deformation leads to the
deformation of the spectrum of the light-cone string oscillators
due to the different masses of different modes and different
fluxes of the NS-NS field in different directions. The dual
description of this solution from the point of view of deformed
CFT states of holographic theory needs further investigation and
will be considered in a forthcoming publication.

{\bf Acknowledgement}

The author would like to thank  R. Poghossian and R. Flume for
collaboration in the early stage of this work, and to thank G.
Savvidy, A. Petkou,  H. J. W. M\"uller-Kirsten and W. R\"uhl for
disscussions. The work was supported by the Alexander von Humboldt
Foundation and in part by the Volkswagen Foundation of Germany and
by INTAS grant \#99-1-590.

\end{document}